\documentclass[12pt]{article}
\usepackage{amsmath, amssymb, physics, geometry}
\usepackage{graphicx}
\usepackage{subcaption}

\title{Optical Emission Spectroscopy Measurements of keV
Apparent Ion Temperatures in Avalanche Energy’s Centrifugal
Mirror Machine}
\author{
\small
M. Affolter$^{*}$, E. C. Hayes$^{*}$, A. Helson, E. McKee, A. Gargone, S. Hepner, and R. Langtry\\
\small Avalanche Energy, 9100 E Marginal Way S, Tukwila, WA 98108, USA\\
\small $^{*}$These authors contributed equally to this work.
}

\date{\small \today}

\begin{document}

\maketitle

\begin{abstract}

Newly formed ions in $E \times B$ devices are rapidly accelerated by strong radial electric fields and execute large cycloidal orbits in the presence of an axial magnetic field. At locations where these orbits intersect, ions originating from different birth radii arrive with substantially different velocities, producing a non-Maxwellian velocity distribution with a large velocity variance. Through Coulomb collisions and collective interactions, this distribution relaxes toward a drifting Maxwellian in the rotating frame. Here, we present the first optical emission spectroscopy (OES) measurements of the line-of-sight–convolved ion-velocity distribution, from which an apparent ion temperature is determined, in Avalanche Energy’s centrifugal mirror machine. High-resolution $H_\alpha$ spectra obtained along five chordal lines-of-sight spanning the plasma radius are analyzed using two complementary models representing limiting cases of the ion dynamics: a collisionless cycloidal model based on the ion-velocity distribution arising from deterministic single-particle orbits, and a rotating Gaussian model based on collisions and collective processes that fully randomize the cycloidal motion into a drifting Maxwellian in the rotating frame. Combined, these approaches bracket the possible degree of velocity-space relaxation and provide a stringent test of the inferred ion energies. Both models reproduce the measured spectra relatively well and yield density-weighted apparent ion temperatures of $1.40\pm0.43$ keV for the rotating Gaussian model and $1.55\pm0.24$ keV for the cycloidal model.  These results provide direct spectroscopic evidence that strong $E \times B$ rotation in a device only a few centimeters in size can generate ion populations with keV energy spreads.
\end{abstract}

\section{Introduction}

Avalanche Energy is pursuing a centrifugal mirror device as a compact neutron source and eventual path to fusion power generation. Similar to other centrifugal mirrors, Avalanche's machine consists of a magnetic mirror field geometry with a perpendicular electric field.  Previous experiments on rotating plasmas confined in a similar device have shown apparent ion temperatures, defined as the mean kinetic energy in the rotating frame, above $20$ keV~\cite{PSP2_Abdrashitov}. At low density or short pulse durations, the ion distribution function of a centrifugal mirror is typically modeled through cycloidal orbits~\cite{Cyc_Ryutova,Cyc_jorgensen}. In contrast, for higher densities and longer pulses, thermal distributions are more valid~\cite{schwartz2024mctrans++}. 

Here, we present experimental measurements of the fast ion population on our compact fusion device through optical emission spectroscopy (OES). The apparent ion temperature of the ions is extracted by fitting this data using both a cycloidal and rotating Gaussian model. These models capture the asymptotic limits of the ion interactions. The cycloidal model is based on non-interacting ions, which is valid for low densities and/or short pulse durations. In contrast, the rotating Gaussian model is the opposite extreme where the ion orbits have been randomized through collisions. Our plasma lies between these limits, and both models yield best-fit density-weighted apparent ion temperatures above $1$ keV.  

The remainder of this paper is organized as follows. Section~\ref{Sec:ExpApp} describes the experimental apparatus, pulsed operating conditions, spectrometer layout, calibration procedure, and the pre-processing steps applied to the measured $H_\alpha$ spectra. Section~\ref{Sec:CycMod} develops the collisionless cycloidal model used to predict the line-of-sight-integrated velocity distribution from single-particle ion orbits, including a benchmark comparison against particle-in-cell simulations. Section~\ref{Sec:RotGaus} presents the rotating Gaussian model, which represents the opposite limit in which collisions and collective processes relax the ion motion into a drifting Maxwellian in the rotating frame. Section~\ref{Sec:Fitting} describes the fitting procedure used to compare both models with the measured spectra and to infer radial profiles of density and apparent ion temperature. Finally, Section~\ref{Sec:Sum} summarizes the inferred apparent ion temperature bounds and discusses their implications for compact centrifugal mirror operation.

\section{Experimental Apparatus}
\label{Sec:ExpApp}
\subsection{Device Geometry}

\begin{figure}
    \centering
    \includegraphics[width=0.8\linewidth]{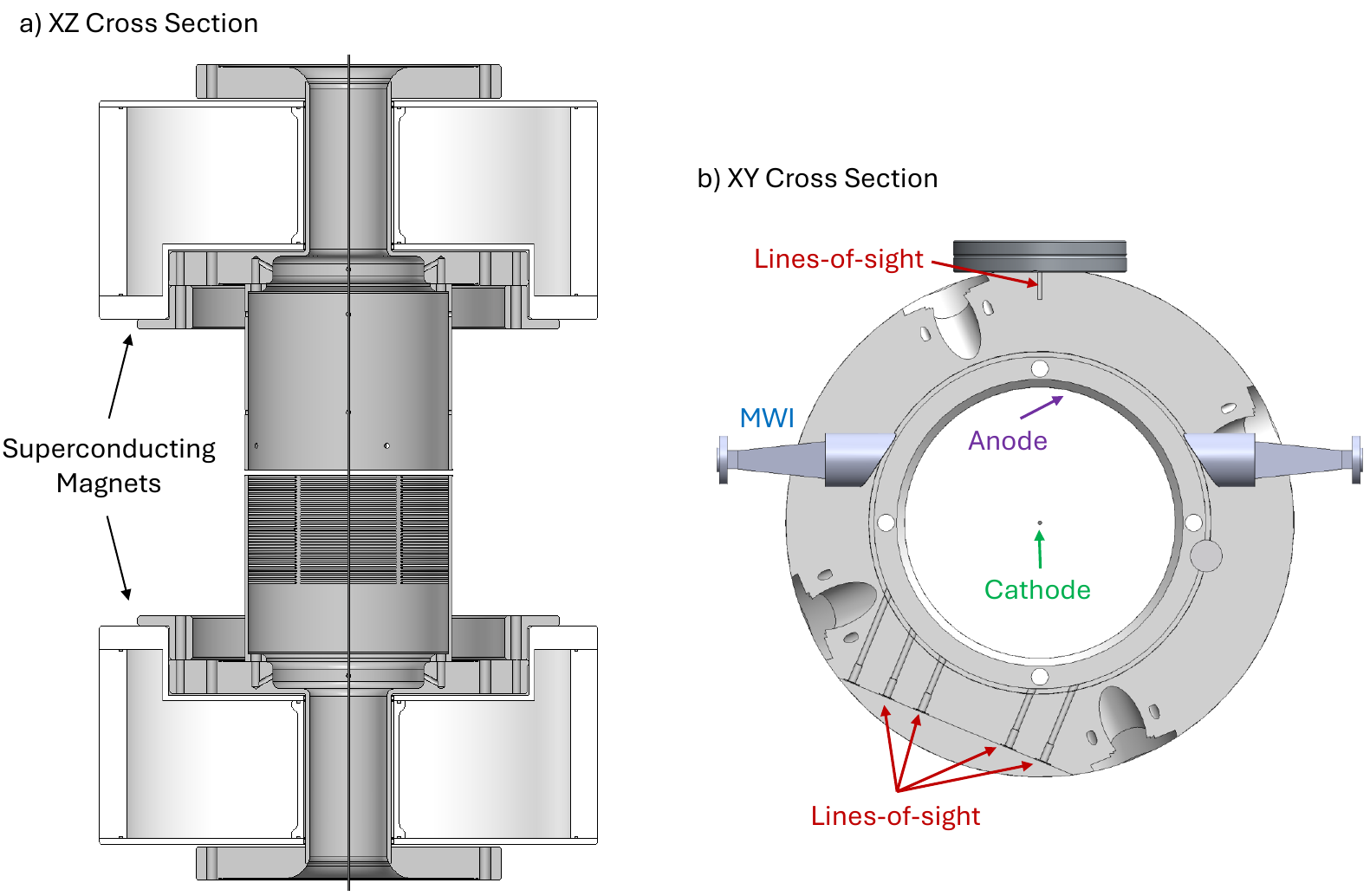}
    \caption{Schematic cross-sectional views of Avalanche Energy's centrifugal mirror machine in both the XZ (a) and XY (b) planes.  Two superconducting magnets form the magnetic mirror field with $B_z=0.45$T on center and a mirror ratio $R_m=10$.  The central cathode is biased negatively relative to the surrounding grounded anode $b\sim6.6$ cm, producing a radial electric field $E_r$ in the presence of an axial magnetic field $B_z$. When the voltage is applied, the background hydrogen gas at pressure $\gtrsim10^{-4}$ Torr breaks down to form the plasma.  The $E\times B$ drift drives azimuthal plasma rotation. Diagnostics include chordal optical LOS, microwave interferometry, and cathode/anode current measurements.}
    \label{fig:LOS}
\end{figure}

Avalanche has been focusing on small radii devices to study plasmas subject to intense electric fields.  Our device, schematically shown in Fig.~\ref{fig:LOS}, consists of a thin central conductor, which is typically negatively biased, surrounded by an outer electrode of radius $b\sim6.6$ cm that is grounded.  When voltage is applied to the central conductor, a predominantly radial electric field $E_r$ is formed.  This device is contained in an axial mirror field with $B_z=0.45$T on center, and a mirror ratio $R_m=10$.  Our main diagnostics are microwave interferometry (MWI), cathode and anode voltage/current measurements, and five chordal lines-of-sight (LOS) used for OES, which is the main focus of this work.

\subsection{Pulsed Operation}

\begin{figure}
    \centering
    \includegraphics[width=0.6\linewidth]{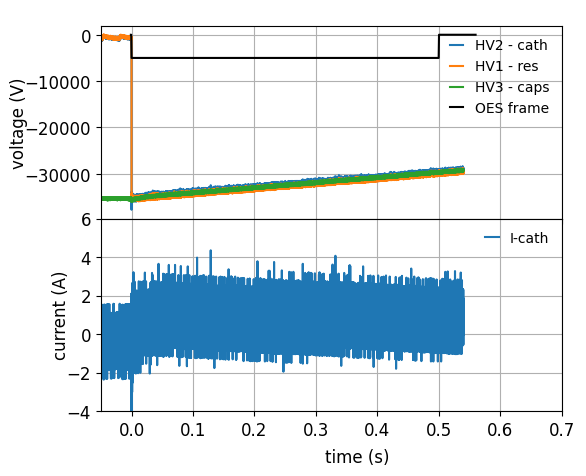}
    \caption{Cathode electrical oscillograms collected during a typical pulsed experiment. The oscillograms show the voltage of the capacitor bank (green) and the voltage measured before and after an inline resistor (orange and blue respectively) leading to the central cathode. The black square wave shows the OES CCD exposure length (500 ms) during the pulsed experiment. The lower figure shows the cathode current calculated from the HV2 and HV1 measurements and the known value of the inline resistor.}
    \label{fig:PulseData}
\end{figure}

Figure~\ref{fig:PulseData} shows typical voltage/current oscillograms generated in the device. The plasma is formed by introducing hydrogen gas at pressures above $10^{-4}$ Torr.  At $t=0$ ms, $-38$ kV is applied to the central conductor through a capacitor bank. The gas breaks down after the cathode has reached peak voltage with an initially higher current draw.  The plasma is sustained by continuously flowing gas over the course of the measurement.

These experiments achieve an electric field of roughly $0.6$ MV/m with $-38$ kV applied to the central cathode. We have developed a high voltage, 6 inch bushing~\cite{borghei2025hammerhead} capable of applying $-300$ kV in this compact volume. As we increase the device radius and magnetic field, these higher potentials will be required.

\subsection{Spectrometer Layout and Calibration}
\label{Sec:DeviceOESDescript}
\begin{figure}
    \centering
    \includegraphics[width=0.9\linewidth]{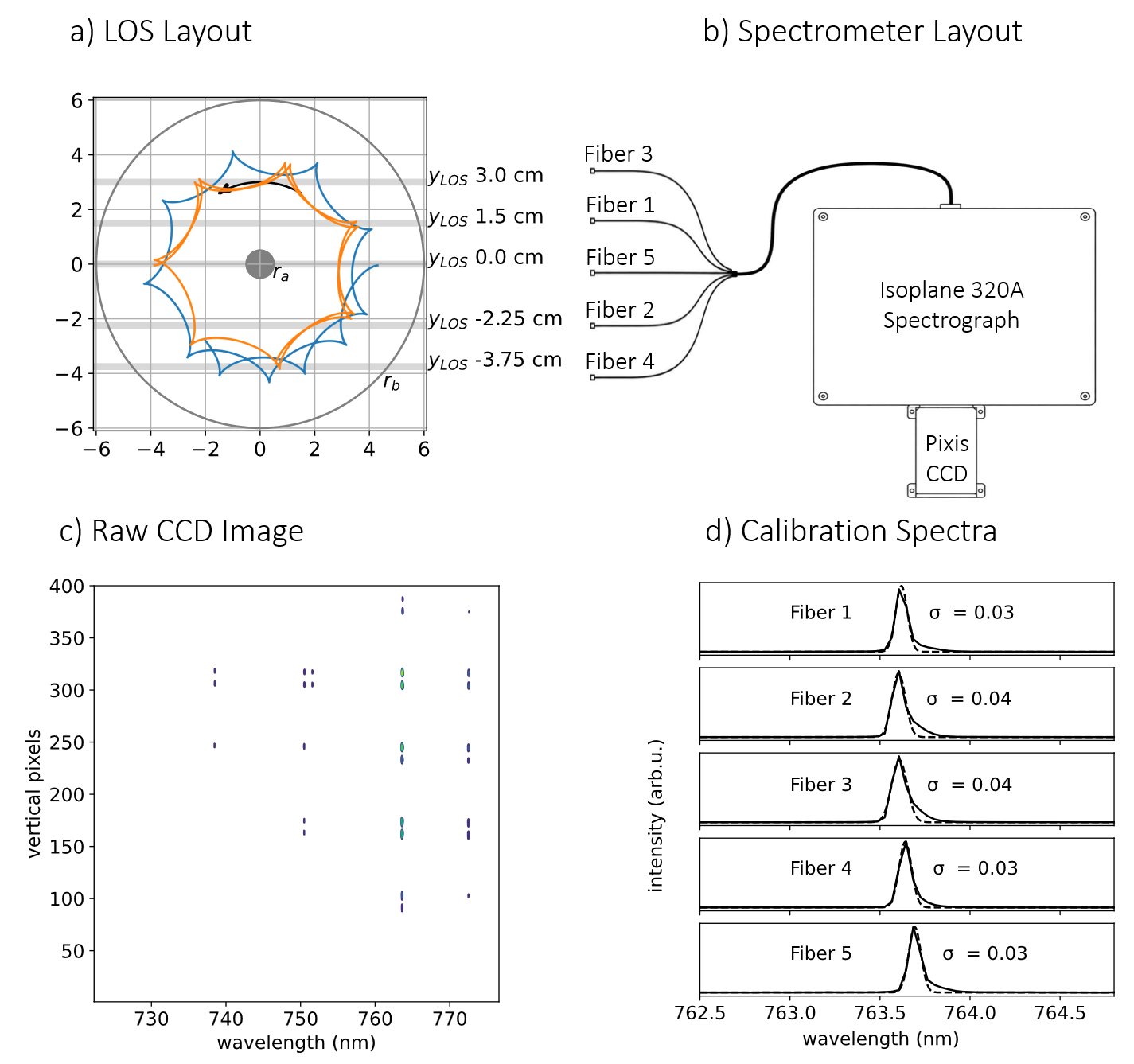}
    \caption{a) A schematic of the Avalanche Energy centrifugal mirror machine including the optical emission LOS at vertical positions $y_{LOS}=(-37.5,-22.5,0.0,15.0,30.0)$ mm. The $E\times B$ drift drives azimuthal plasma rotation along the guiding center (black arrow) with cycloidal orbits (blue and orange). Each viewing chord subtends a finite diameter that increases from approximately $3$mm to $8$mm across the plasma and is coupled by b) a multileg optical fiber bundle to the spectrometer entrance slit. 
    c) The raw CCD image shows the emission from a low-temperature spectral lamp used to measure the spectrometer line-spread function. d) Each LOS is integrated vertically to produce the spectra (solid curves). The one-standard-deviation widths ($\sigma_{inst}$) of Gaussian fits to the calibration lamp peaks are indicated in each panel.}
    \label{fig:Spec}
\end{figure}

The focus of this work is to diagnose the apparent ion temperature of the plasma through the use of OES of the excited charge exchange $H$ neutrals.  Figure~\ref{fig:Spec} shows the five LOS located at positions $y_{LOS}=$ $(-37.5$, $-22.5$, $0.0$, $15.0$, $30.0)$ mm on each side of the central cathode.  The plasma at each LOS is viewed through a $2.5$ mm hole in the anode. In-vacuum plano-convex lenses (FL = $6$ mm, OD = $3$ mm) are placed at the same x-coordinate outside the anode and focus plasma emitted light onto an air-to-vacuum fiber optic feedthrough. The optical path diameter of each LOS slightly diverges across the plasma ranging from $w=3$ mm to $w=8$ mm. The optical feedthroughs are coupled to a multileg fiber bundle arranged vertically in a single $10$ mm OD ferrule and coupled to the entrance slit of the spectrometer. This allows for simultaneous OES measurement along the different LOS.

The spectrometer is a $0.32$ m focal length spectrograph (Isoplane 320A, Teledyne Princeton) equipped with a three-grating turret that images onto a CCD (Pixis, Teledyne Princeton). For these measurements, we use a $1200$g/mm grating and optimize the variable entrance slit of the spectrometer for high wavelength resolution while maximizing the light intensity. Wavelength, resolution, and relative intensity calibrations are accomplished with low temperature pencil lamps (Ar, HgAr, Ne Newport MKS). The onboard wavelength calibration routine and dispersion calculator provided by Teledyne-Princeton perform the pixel-to-wavelength calibration, which is confirmed within the $H_\alpha$ region of interest to be within 0.2 nm of the NIST emission lines for argon. The data were corrected for this residual $\sim0.2$ nm offset with a systematic shift. Gaussian fits to the OES instrument response indicate a broadening of 0.04 nm (sigma) or $\sim0.1$ nm (FWHM) as shown in Fig.~\ref{fig:Spec}, which established the instrument response function (IRF). Intensity calibration of spectra along each LOS is performed for both the air-to-vacuum optical feedthroughs and for the multileg fiber optic to the sensor.  The air-to-vacuum optical feedthrough is intensity calibrated by placing the lamp at fixed locations across each LOS while the chamber is open for maintenance.  All but the $y_{LOS}=0.0$ mm have undergone this intensity calibration.  The multileg fiber optic to sensor calibration is done by simultaneously mounting each leg in front of the lamp and imaging all legs onto the sensor. Variations in power to each leg are accounted for and the measurement is done for each exposure utilized in the data collection. From this, we determine a scaling factor for each LOS that is applied to the data before fitting.

\subsection{OES Measurements}

Figure~\ref{fig:RawCorr} shows OES spectra collected over a time period of $500$ms during a pulsed experiment. OES is used to detect both the cold $H$-neutral population (sharp peak centered at 656.3 nm) and the fast $H$ neutral population (low intensity wings around the cold peak). The fast neutral population is generated through charge exchange of fast ions on the elevated background of neutral $H_2$. LOS on opposing sides of the cathode (negative/positive y-values) display higher intensities on the blue/red side of the $H_\alpha$ line.  These spectra, which are LOS integrated measurements, show that the plasma is rotating.  The Doppler broadening for the LOS pointed at the cathode ($y_{LOS}=0$) is symmetric since it measures no net plasma rotation except for rotational broadening captured within the finite LOS width in $y$. For reference, a $1$ keV hydrogen ion moving parallel to the LOS is shifted by about $1$ nm.

\subsection{OES Data Corrections}

\begin{figure}
    \centering
    \includegraphics[width=0.89\linewidth]{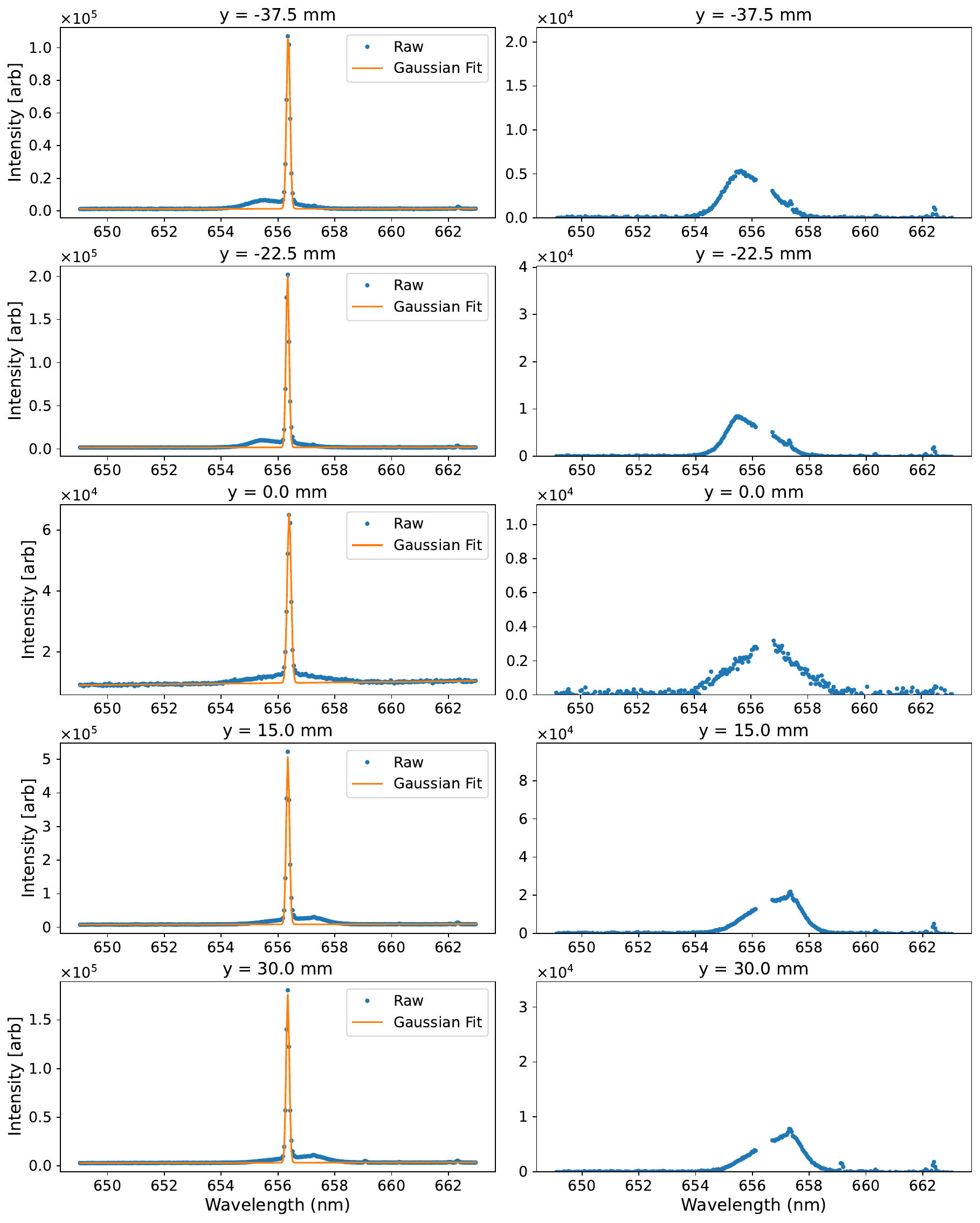}
    \caption{Measured $H_\alpha$ emission spectra along five different LOS $y_{LOS}=(-37.5,-22.5,0.0,\\15.0,30.0)$ mm. Left column: raw optical spectra containing the dominant narrow $H_\alpha$ emission from cold neutral hydrogen superimposed on a small background offset. The cold component is fitted with a Gaussian profile (orange curve). Right column: spectra after background subtraction and slight wavelength calibration, with the cold neutral component shifted to the rest wavelength of $656.28$ nm. The spectral region dominated by the cold neutral line (orange points) is excluded, leaving the broadened wings (blue points) associated with fast neutrals produced by charge exchange.}
    \label{fig:RawCorr}
\end{figure}

Before each pulsed experiment, a background image is collected with the camera shutter closed. This is subtracted from the plasma image to remove thermal charge accumulation on the CCD. We remove residual background offset, attributed to light leakage into the vacuum chamber, by applying a linear background fit below and above the $H_\alpha$ spectrum and subtract this offset for each LOS.  We then fit a Gaussian to the cold neutral emission line, defined by
\begin{equation}
I_{\text{neutral}}(\lambda) = A \exp\left[-\frac{(\lambda - \lambda_0)^2}{2\sigma^2}\right],
\end{equation}
where $A$ is an arbitrary amplitude, $\lambda$ is wavelength,  $\lambda_0$ represents the peak wavelength, and $\sigma$ is the variance.  We remove a region around this peak from the dataset before model fitting to ensure the fully reduced spectrum is entirely due to the fast-neutral population. Figure~\ref{fig:RawCorr} shows the experimental spectra before (Left) and after (Right) the corrections.

\label{Sec:OESDataCorr}

\subsection{LOS Integrated Measurement}
\label{Sec:LOS_Con}
Because the plasma exhibits bulk rotation, the measured spectral broadening shown in Fig.~\ref{fig:RawCorr} contains contributions from the finite volume of the viewing chord.  This means that the LOS convolution must be considered when determining the apparent ion temperature.  At each point $(x,y)$ along the LOS, light with a wavelength $\lambda$ and intensity $I(\lambda,x,y)$ can be detected.  The measured spectra are therefore described by the integral
\begin{equation}
I_{\mathrm{lab}}(\lambda,y_{\mathrm{LOS}})=
\int_{\mathrm{LOS}} I(\lambda,x,y)\, W(x,y)\, dx\,dy
\label{Eq:I_Conv}
\end{equation}
over the viewing chord, where $W(x,y)$ accounts for fall off in the light collection efficiency with distance $x$ and the width of the collection optics in $y$.

We are directly measuring the spectra of excited charge exchange neutrals, rather than the ion distribution. To extract the apparent ion temperature from this signal, we assume that newly created neutrals inherit the ion velocity distribution $f(v,r)$ and that effects such as energy-dependent charge exchange cross sections do not significantly distort the distribution.  With this assumption, the local intensity spectra are expressed as
\begin{equation}
I(\lambda,x,y)\propto
n(r)\int f(v_x,r)\delta\left[\lambda -\lambda_0\left(1+\frac{v_x}{c}\right)\right]dv_x,
\label{Eq:I_local}
\end{equation}
where we have taken the LOS direction to define the $x$-axis, which means that the Doppler broadening arises from the velocity component $v_x$.  With Eqs.~\ref{Eq:I_Conv} and~\ref{Eq:I_local}, we fit the measured spectra to obtain the ion velocity distribution.  In the next section, we introduce the cycloidal and rotating Gaussian models of the velocity distribution function.

But, first, let us discuss the assumption that the charge exchange neutrals inherit the ion velocity distribution.  The velocity distribution function of the charge exchange neutrals is a charge exchange reaction-rate-weighted representation of the ion velocity distribution, where the weighting is determined by the velocity dependence of the reaction rate coefficient, $\sigma(v)v$. Since this factor depends on the speed of the ions, this approximation holds well for ion populations where the bulk velocity is much greater than the thermal velocities.  The cross section $\sigma$ for fast ions to charge exchange and produce excited neutrals~\cite{Phelps_CS} increases by over an order-of-magnitude from $100$ eV to $1$ keV.  For a plasma without bulk flow, this would narrow the inferred ion velocity distribution compared to the measured neutral distribution since more low energy ions are required to produce that intensity of light.  However, for our rotating plasmas, the energy dependence of the cross section will skew the distribution since the bulk of our ions are moving with considerable energy compared to the neutral background.  The exact impact that this has on the apparent ion temperature measurements will be incorporated into future modeling.

\section{Cycloidal Model}
\label{Sec:CycMod}

In this section, the single-particle orbit-averaged model based on the non-interacting cycloidal velocity distribution is discussed to determine the density-weighted average apparent ion temperature.  Previously developed theory~\cite{Cyc_jorgensen} is based on a parabolic Larmor velocity distributions with orbital radii approximated from a parallel plate capacitor to determine the LOS integrated velocity distribution perpendicular to the magnetic field.  We expand upon this theory by direct construction of the velocity distribution function $f(v,r)$ from conserved single-particle invariants without assuming an ad hoc velocity distribution.

\subsection{Single-Particle Ion Orbits}
To simplify the problem, a coaxial cylindrical geometry with inner radius $a$ at potential $-V$ and outer radius $b$ at ground is used. Space charge is neglected assuming a low-density plasma, which means that the electrostatic potential is just the vacuum potential
\begin{equation}
\Phi(r) = -V \frac{\ln(r/a)}{\ln(b/a)}.
\end{equation}
A uniform axial magnetic field $\mathbf{B} = B_z \hat{z}$ is assumed.  As is typically observed in the experiments, the ionization occurs after the electric field is established, which means that the fields can be treated as static.  

In these time-independent and azimuthally symmetric fields, both the energy
\begin{equation}
E = \frac{1}{2} m (v_r^2 + v_\theta^2) + q \Phi(r)
\label{Eq:Energy}
\end{equation}
and canonical angular momentum
\begin{equation}
P_\theta = m r v_\theta + \frac{q B_z}{2} r^2
\end{equation}
are conserved.  For ions born at rest at radius $r_0$,
\begin{equation}
P_\theta = \frac{q B_z}{2} r_0^2,
\end{equation}
and the azimuthal velocity at the observation radius $r$ is
\begin{equation}
v_\theta(r, r_0) = \frac{q B_z}{2m} \left( \frac{r_0^2}{r} - r \right).
\end{equation}  
Using conservation of energy $E(r)=E(r_0)$ and Eq.~\ref{Eq:Energy} with the ions born at rest at $r_0$, the radial velocity is expressed as
\begin{equation}
v_r^2 = F(r, r_0)= \frac{2q}{m} [\Phi(r_0) - \Phi(r)] - v_\theta^2(r, r_0),
\end{equation}
where $\Phi(r_0) - \Phi(r)=V[\ln(r_0/r)/\ln(b/a)]$ assuming a coaxial device.  The orbital turning points $(r_-, r_+)$ for a given $r_0$ are determined numerically by solving for the roots of $F(r, r_0)=0$.  An orbit is considered bound in this system if the lower turning point is larger than the cathode radius $r_->a$.  The oscillation period $T_r(r_0)$ of these bound orbits is determined through numerical integration of
\begin{equation}
T(r_0)=2\int_{r_{-}}^{r_{+}} \frac{dr}{v_r(r,r_0)},
\label{Eq:Period}
\end{equation}
where $v_r(r,r_0)=\sqrt {F(r, r_0)}$.

\subsection{Orbit Averaged Velocity Distribution}

As ions are uniformly distributed in time along their bound trajectories, averaging the velocity distribution over an orbital period is valid. The resulting distribution is constructed by conserving particle number and weighting by time spent near each radius. The radial velocity distribution is then
\begin{equation}
f(v_r, r) = \int_a^b dr_0 \frac{r_0 n_0(r_0)}{r T_r(r_0) u} [\delta(v_r - u) + \delta(v_r + u)],
\end{equation}
where $u = \sqrt{F(r, r_0)}$ and $T_r(r_0)$ is the radial period calculated previously.  At each observation radius $r$, there is a probability of finding $2\pi n_0(r_0)r_0dr_0$ particles with velocity $\pm v_r$ (descending versus ascending).  The probability of finding those particles at that $r$ depends on the duration of the orbit spent at that radius $dr/(v(r,r_0)T(r_0))$.  The azimuthal velocity distribution follows the same form
\begin{equation}
g(v_\theta, r) = \int dr_0 \frac{2 r_0 n_0(r_0)}{r T_r(r_0) u} \delta(v_\theta - v_\theta(r, r_0)),
\end{equation}
but at each radius there is only one azimuthal velocity.

To construct the ion velocity distributions $f(v_x,r)$ representing these measured spectra, we note that
\begin{equation}
v_x = \frac{x}{r} v_r - \frac{y}{r} v_\theta,
\end{equation}
for a point $(x,y)$ in the plasma, where $r = \sqrt{x^2 + y^2}$.  The terms $x/r$ and $y/r$ determine the $x$ component of $v_r$ and $v_\theta$, respectively.  At each observation radius $r$, an ion orbit born at $r_0$ contributes two radial velocities $\pm v_r$ for each $v_{\theta}$ and each $(r, r_0)$ pair, which produces two contributions
\begin{equation}
v_x = \frac{x}{r} (\pm v_r) - \frac{y}{r} v_\theta(r, r_0).
\end{equation}
Therefore,
\begin{equation}
f(v_x, r) = \int dr_0 \, \frac{r_0 n_0(r_0)}{r T_r(r_0) u}
\sum_{\pm} \delta(v_x - v_x^{\pm}),
\end{equation}
and the LOS-convolved intensity spectrum is determined from Eqs.~\ref{Eq:I_Conv} and~\ref{Eq:I_local}.

\subsection{Comparison to Particle-in-Cell (PIC) Simulation}
\label{Sec:WarpXBench}
WarpX is an open-source PIC code that is broadly utilized at Avalanche Energy to perform simulation studies of our various plasmas. To benchmark the theory presented here, we disabled several of the code's multi-physics models, including the space charge deposition to the grid and the pair-wise Monte Carlo Coulomb collision model.  Using the code's Boris particle pusher as an orbit integrator, we ran a kinetic 3D coaxial simulation with $V=-5$ kV, $B=0.45$ T, $a=1$ cm, $b=10$ cm, and a uniform radial initial ionization profile $n_0(r)$ (these parameters were chosen for ease of simulation).  We ran the simulation for $200$ cyclotron periods.  In the first $10$ periods, ions are randomly initialized in the volume with a temperature of $300$ K, such that the orbits will be phase mixed.  Following this initial loading period, we allow for $10$ more cyclotron periods to occur before the moments are extracted.

\begin{figure}
    \centering
    \includegraphics[width=0.65\linewidth]{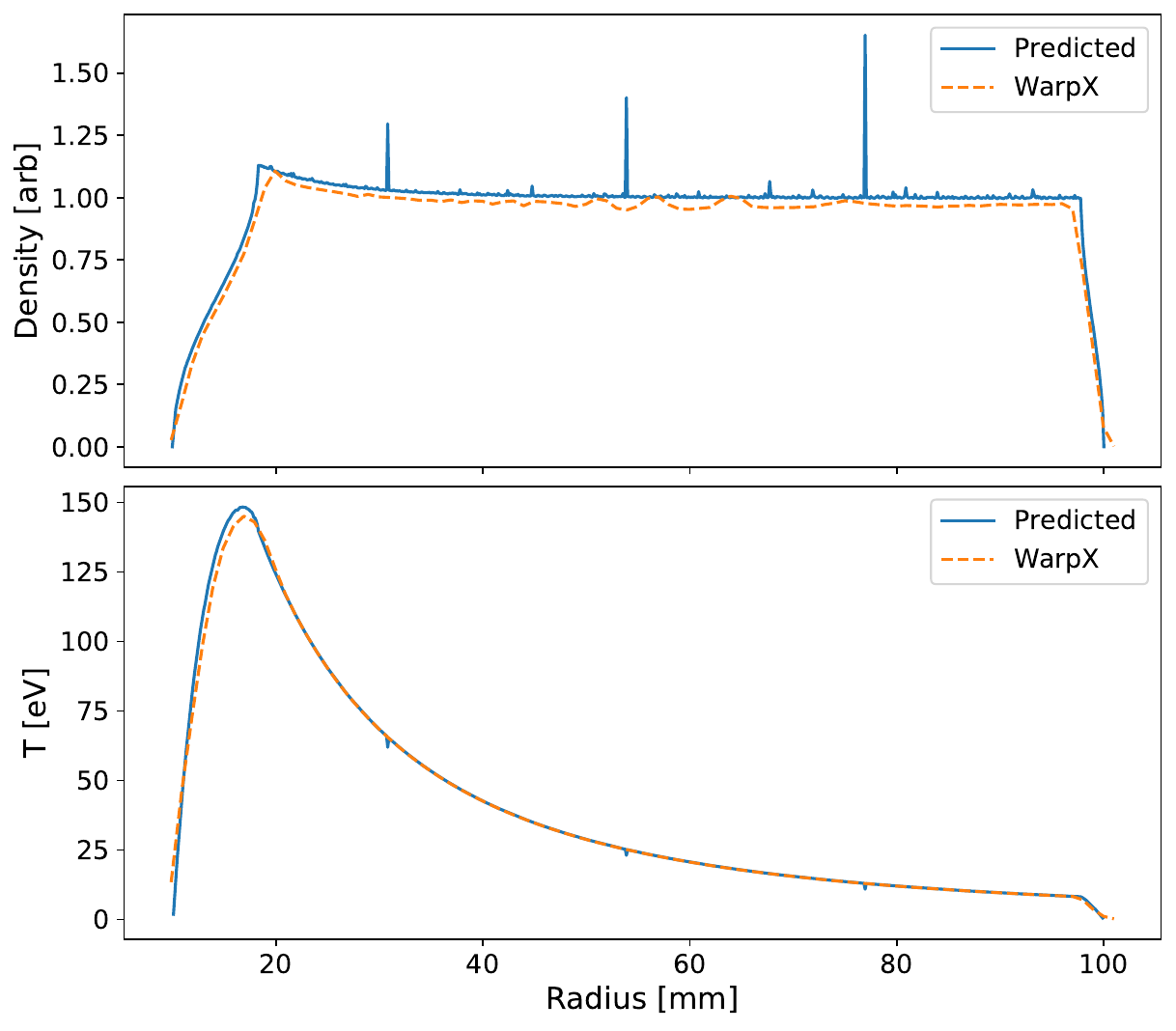}
    \caption{Benchmark comparison between the analytic cycloidal model and WarpX PIC simulations for $V = -5$ kV, $B_z = 0.45$ T, $a = 1$ cm, $b = 10$ cm, and a uniform ionization profile $n_0(r)$. Shown are the resulting radial density and apparent ion temperature profiles. Very importantly, the close agreement validates the orbit model and confirms that large apparent ion temperatures arise naturally from collisionless cycloidal motion.}
    \label{fig:Prof}
\end{figure}

Figure~\ref{fig:Prof} shows the excellent agreement between these PIC simulations (orange curves) and modeled (blue curves) density and apparent ion temperature profiles. WarpX includes a particle temperature diagnostic that is used to calculate the momentum variance of each species on a per-cell basis. This allows for the determination of the apparent ion temperature in the same way defined in the previous section.  However, by default, the code calculates the ion total temperature in the three degrees of freedom.  For these simulations, $T_z=0$ and a lower temperature is found in WarpX.  For direct comparison, we changed the ion temperature calculated in our theoretical model to reflect this difference.

\begin{figure}
    \centering
    \includegraphics[width=0.95\linewidth]{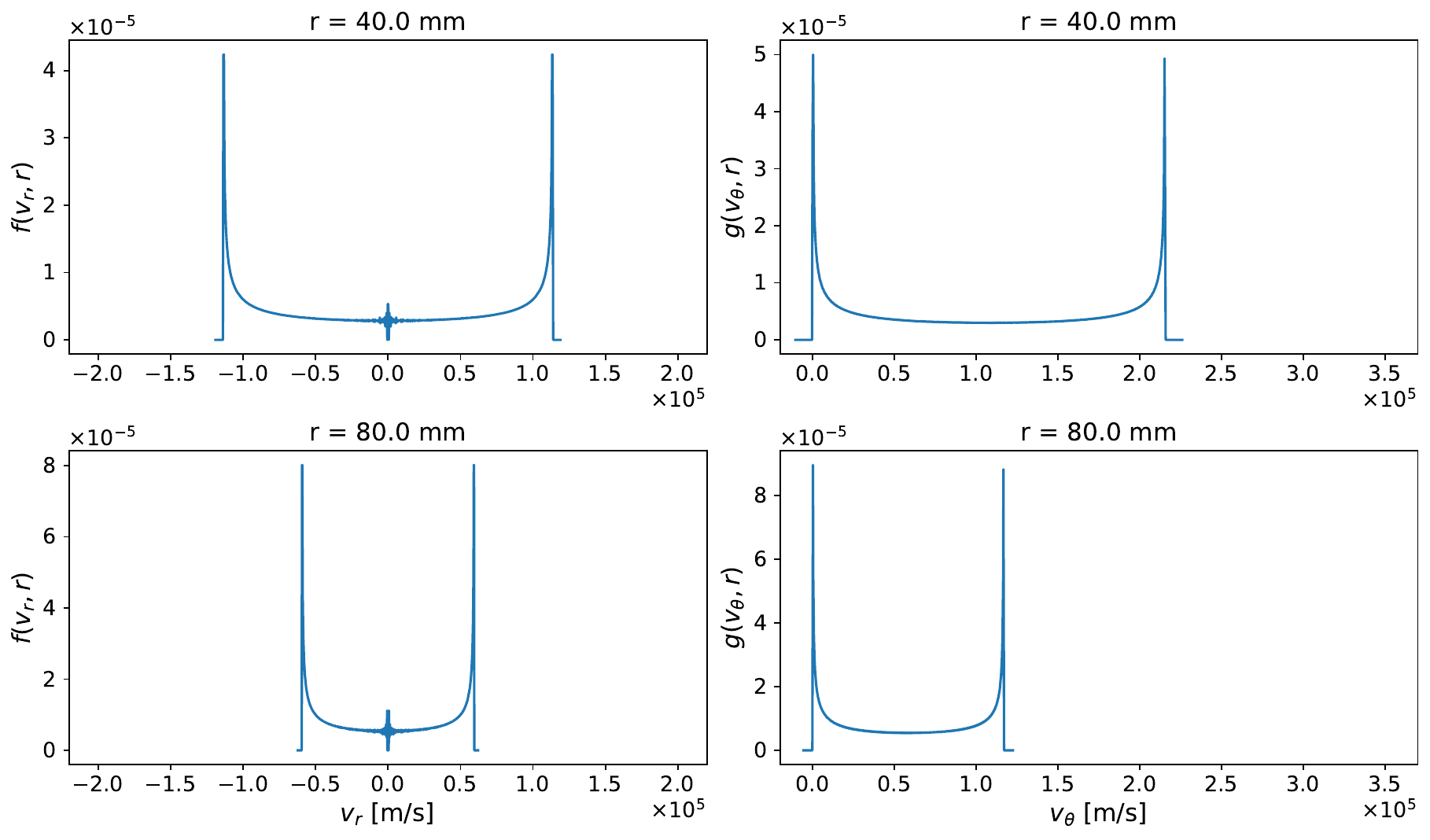}
    \caption{Calculated fast-ion velocity distributions in radial–azimuthal velocity space for the collisionless
    cycloidal model with $V=-5$ kV, $B=0.45$ T, $a=1$ cm, and $b=10$ cm. Distributions are shown at two radial locations to illustrate the multi-valued velocity structure produced when ions born at different radii follow cycloidal orbits that intersect at a common position. The large velocity-space variance generated by orbit crossing leads to substantial Doppler broadening. Numerical noise near $v_r = 0$ is a discretization artifact arising from the radial step size. This parameter set is chosen for benchmarking against WarpX PIC simulations and does not correspond to the experimental geometry.}
    \label{fig:Dist}
\end{figure}

Figure~\ref{fig:Dist} shows the calculated radial and azimuthal velocity distributions at two radial locations using the geometry and voltage from the benchmark comparison. We see that the radial distribution is symmetric about $v_r=0$ as expected with velocity variance that decreases with increasing $r$.  In contrast, the azimuthal velocity distribution is centered around a non-zero velocity that decreases with increasing radius.  This is expected from the decreasing $E\times B$ drift velocity with $r$.

\subsection{Model Fitting with Discrete Ionization Radii}
To efficiently evaluate the model applied to the experimental spectra, we discretize the ionization profile $n_0(r)$ into a set of radial source slices,
\begin{equation}
n_0(r) \approx \sum_k w_k \, \chi_k(r),
\end{equation}
where $\chi_k(r)$ represents a narrow radial shell centered at $r_0^k$, and $w_k$ is its weight.  Because the model is linear in $n_0(r)$, the LOS spectra can be written as
\begin{equation}
I_j(\lambda) = \sum_k w_k B_{j,k}(\lambda),
\end{equation}
where $j$ indexes the LOS, $k$ indexes the radial source slice, and $B_{j,k}(\lambda)$ is the LOS-integrated spectrum produced by ions born only at $r_0^k$.  Each basis function $B_{j,k}(\lambda)$ is computed once by performing the LOS integration numerically with $n_0(r) = \delta(r - r_0^k)$.  The inputs used to determine these basis functions are the measured cathode voltage $V=-33\pm5$kV, the average magnetic field at the LOS position $B_z=0.51$T, the cathode radius $a=0.79$mm, and anode radius $b=66$mm.

To reduce the number of fitting parameters, the number of basis functions is chosen 
to be much smaller than the resolution used in the orbit calculation.  The orbit dynamics are computed on a fine grid in $r_0$ and interpolated smoothly, 
but the fitting uses a reduced set of $N_{\text{basis}}$ radial slices
\begin{equation}
k = 1, \dots, N_{\text{basis}}.
\end{equation}
Typically, $N_{\text{basis}}=30$.  Other values have been chosen for $N_{\text{basis}}$ with little impact to the overall conclusion.

The modeled spectra are constructed as
\begin{equation}
I_j^{\text{model}}(\lambda) = s_j \sum_k w_k B_{j,k}(\lambda),
\end{equation}
where $s_j$ are per-LOS multiplicative scale factors accounting for experimental variations in collection efficiency.  The theoretical spectra are further convolved with the IRF (discussed in Section~\ref{Sec:DeviceOESDescript}) to model the finite spectrometer resolution.  From the fit of this model to the experimental data shown in Figure~\ref{fig:RawCorr}, we determine the ionization profile of the plasma, which is then used to directly determine the density and apparent ion temperature profiles.  This fitting procedure is discussed in Sec.~\ref{Sec:Fitting}.

\section{Rotating Gaussian Model}
\label{Sec:RotGaus}
In the opposite limit of the cycloidal model, the ion orbits randomize through ion-ion collisions or ion-neutral collisions and the velocity distribution is described by a rotating Gaussian.  In this section, we develop this LOS integrated rotating Gaussian model.  As we only measure the $x$ velocity component of the Doppler broadening with OES, we can describe the velocity distribution at each radius $r$ as
\begin{equation}
f(v_x, r) =
\frac{1}{\sqrt{2\pi \sigma_x^2(r)}}
\exp\left[
-\frac{(v_x - v_{rot,x}(r))^2}{2\sigma_x^2(r)}
\right],
\end{equation}
where $v_{\mathrm{rot,x}}(r) = -\omega_{E\times B}(r)\,y$ is the x-component of the $E\times B$ rotation velocity and $\sigma_x^2(r)$ is the x-component velocity variance in the rotating frame.  We again neglect space charge and assume that the rotational velocity is determined by the vacuum $E \times B$ profile.  For a coaxial geometry, the electric field is given by
\begin{equation}\label{eqn:electric_field}
E_r = \frac{V}{r \ln(b/a)},
\end{equation}
and the velocity as
\begin{equation}
v_{E\times B}(r) = \frac{1}{r}\frac{V}{\ln(b/a)B(r)}.
\end{equation}

Again, using Eqs.~\ref{Eq:I_Conv} and~\ref{Eq:I_local}, we construct the LOS integrated intensity spectrum \\$I(v_x, y_{LOS})$.  For this model, we assume that the ion lifetime is longer than the axial bounce time, which means that the plasma density is restricted to the region between the inner and outer magnetic field limiters located at $10$mm and $57$mm, respectively.  A simple piecewise linear density profile is assumed between these limits with the peak location variable. This peaked density profile is supported by the fact that the cold $H_\alpha$ intensity decreases from a maximum at $y_{LOS}=15.0$mm as the LOS position is increased for the calibrated LOS shown in Figure~\ref{fig:RawCorr}.  Fitting this model to the experimental data, the apparent ion temperature profile in the rotating frame is determined.

\section{Determining the Apparent Ion Temperature}
\label{Sec:Fitting}
We have presented two models for the ion velocity distribution function, which represent asymptotic limits of the ion collision rate.  Our plasma lies between these limits.  However, by fitting the experimental data using both models, bounds on the apparent ion temperature are determined.

We define the apparent ion temperature of each degree of freedom to be
\begin{equation}
\frac{1}{2}k_BT_{app} = \frac{1}{2}m\sigma_v^2,
\end{equation}
where $\sigma_v^2=\langle v^2\rangle-\langle v\rangle^2$ is the velocity variance of a component with moments
\begin{equation}
n(r) = \int_{-\infty}^{\infty}f(r,v)dv,
\end{equation}
\begin{equation}
\langle v\rangle=\frac{1}{n}\int_{-\infty}^{\infty}vf(r,v)dv,
\end{equation}
and
\begin{equation}
\langle v^2\rangle=\frac{1}{n}\int_{-\infty}^{\infty}v^2f(r,v)dv,
\end{equation}
where $n$ is the average ion density.  In the rotating frame, the mean velocity vanishes $\langle v\rangle=0$, and the apparent ion temperature is determined directly by the velocity variance $k_BT_{app}/2=m\langle v^2\rangle/2$. To prescribe a single temperature to the system, we define the density-weighted average apparent ion temperature as
\begin{equation}
\langle T_{\mathrm{app}} \rangle = \frac{\int_a^b r n(r) T_{\mathrm{app}}(r) dr}{\int_a^b r n(r) dr}.
\end{equation}

To extract the apparent ion temperature, the same fitting procedure is used for both models.  We use a reduced $\chi^2$ approach that minimizes the residual between the modeled and experimental spectra for all lines-of-sight $j$ simultaneously.  The reduced $\chi^2$ is 
\begin{equation}
\chi^2 = \frac{1}{N-p}\sum_j \sum_{\lambda \in \mathrm{data}}
\left( \frac{I_j^{\mathrm{model}}(\lambda) - I_j^{\mathrm{data}}(\lambda)}{\sigma_j(\lambda)} \right)^2,
\end{equation}
where $N$ is the number of data points used in the fit and $p$ is the number of fitting parameters.  Each LOS is allowed a multiplicative scale factor to account for variations in collection efficiency.  For simplicity, we assume that the LOS across the plasma are collimated in both models.  We account for the diverging LOS width by running both models with the LOS diameter extrema.

\subsection{The Application of the Cycloidal Model to Data}
\label{sct:cycloidal-model-fitting}
\begin{figure}
    \centering

    \begin{subfigure}{0.495\textwidth}
        \centering
        \includegraphics[width=\linewidth]{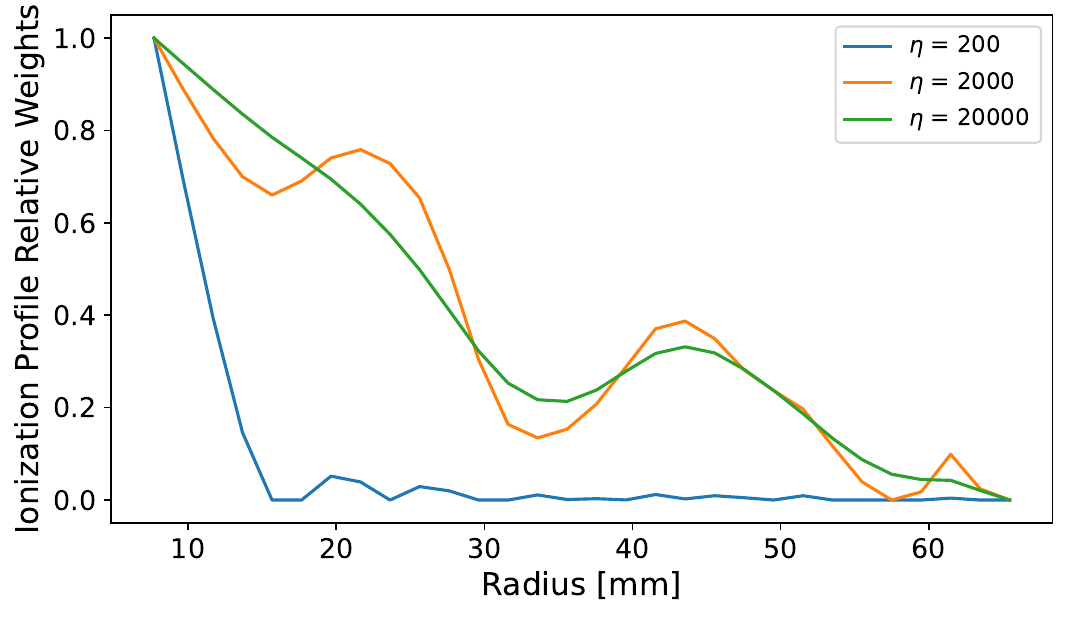}
    \end{subfigure}
    \hfill
    \begin{subfigure}{0.495\textwidth}
        \centering
        \includegraphics[width=\linewidth]{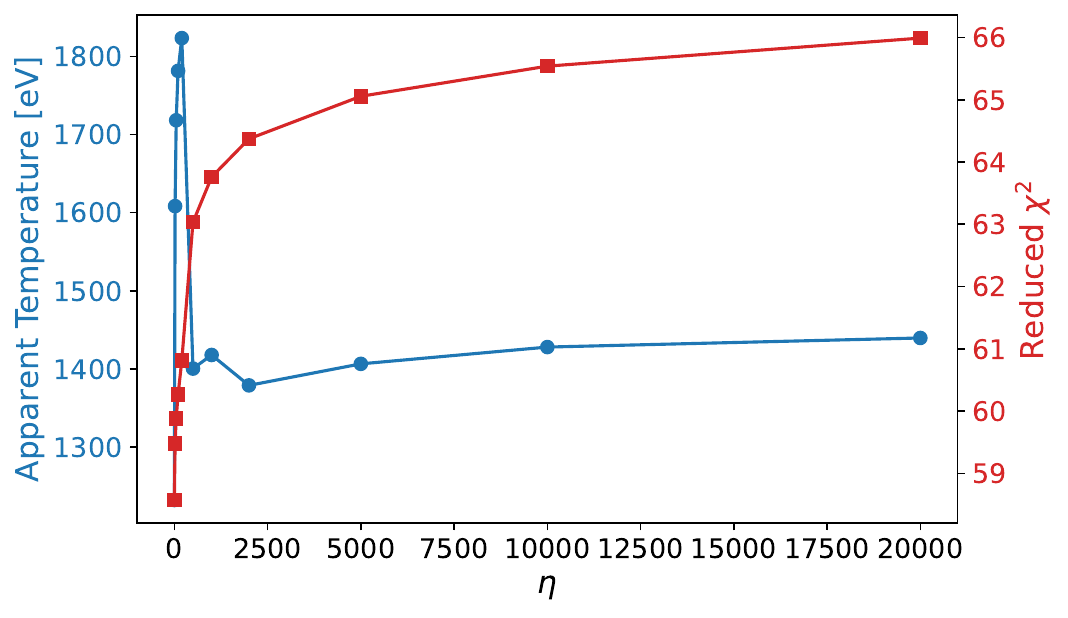}
    \end{subfigure}

    \caption{(Left) Ionization profiles for different smoothing parameters determined through fitting the experimental data with the cycloidal model with a $V=-33$ kV.  As the value of the smoothness parameter $\eta$ is increased, large variations in the predicted ionization profile are more heavily penalized in the fitting.  The blue curve ($\eta=200$) shows the unphysical result of discrete ionization bands when too little smoothing is applied.  (Right)  As the smoothing parameter is increased, the deviation between the model and experimental data (represented by the reduced $\chi^2$) becomes worse.  We choose an $\eta=2000$ to enable variations in the ionization profile while preventing large oscillations.  The choice of $\eta$ provides a lower estimate of the density-weighted apparent ion temperature (blue curve). }
    \label{fig:IonProf}
\end{figure}

For the cycloidal model, the ionization profile is determined through the fitting process given the measured cathode voltage, average magnetic field, and geometry.  To stabilize the solution, a smoothness regularization is applied to the ionization profile weights
\begin{equation}
\eta \sum_k \left( w_{k+1} - 2 w_k + w_{k-1} \right)^2,
\end{equation}
where $\eta$ represents the strength of the smoothing.  Figure~\ref{fig:IonProf} shows the fitted ionization profile (Left) for various smoothing parameters.  Without smoothing, the fit produces a unphysical result of discrete ionization bands.  As the smoothing parameter is increased, the ionization profile is initially strongly peaked near the cathode and then has gradual oscillations as $\eta$ is increased further.

Figure~\ref{fig:IonProf} (Right) shows how the deviation between data and fit (reduced $\chi^2$) and density-weighted apparent ion temperature vary with $\eta$.  We choose the smoothing parameter $\eta=2000$ that enables variations in the ionization profile but reduces large unphysical oscillations.  This choice does not strongly change the resulting apparent ion temperature.  The apparent ion temperature in the rotating frame is calculated from this ionization profile via the process described in Sec.~\ref{Sec:Fitting}.

\begin{figure}
    \centering
    \includegraphics[width=0.95\linewidth]{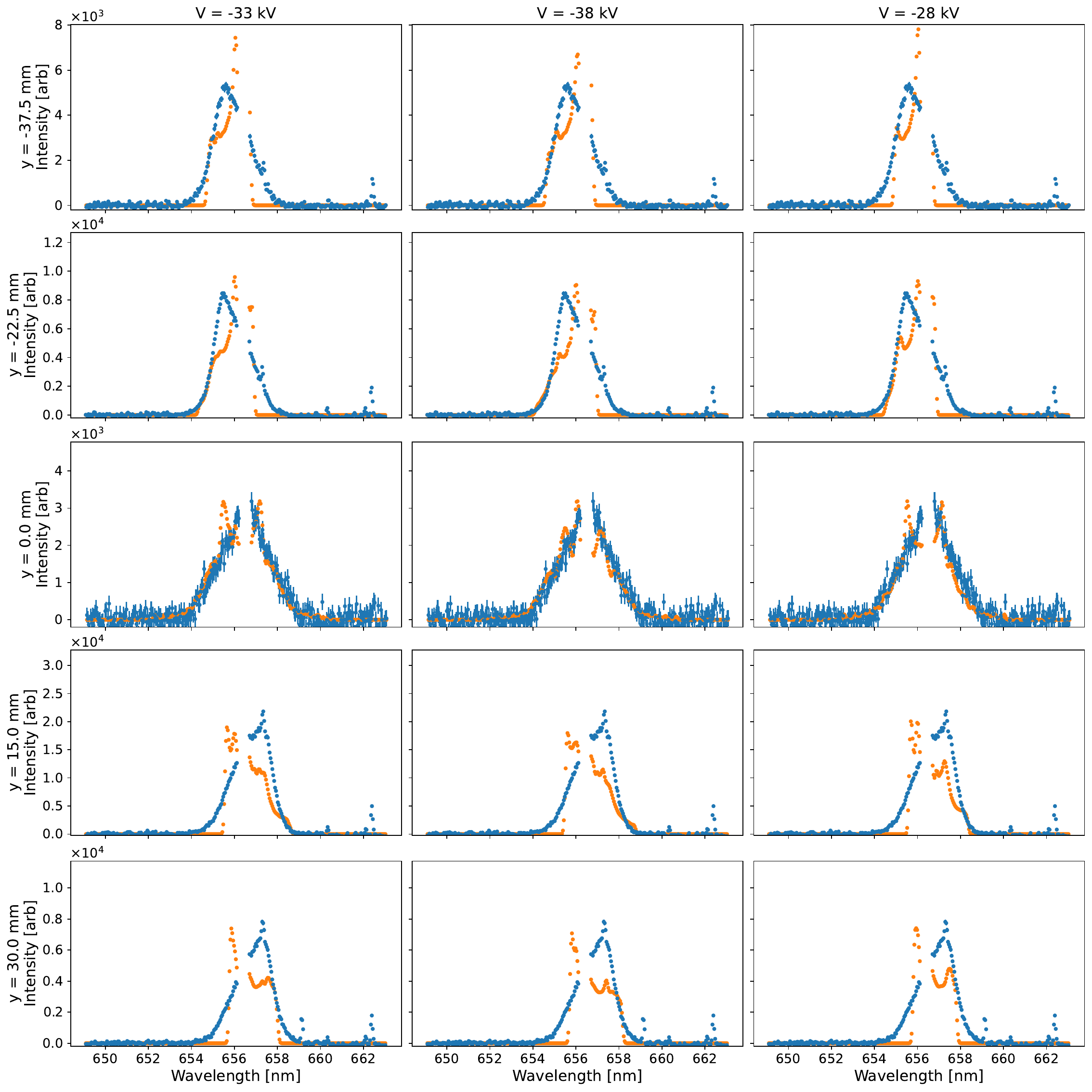}
    \caption{Measured $H_\alpha$ spectra (blue points) and best-fit model predictions (orange points) for the five OES viewing chords using the collisionless cycloidal model. The three different columns show separate fits using the average measured voltage $-V=33\pm5$ kV and voltage bounds.  The fits generally have sharper features than the experimental data most likely resulting from the presence of collisions, but they do capture the width of the distribution fairly well.  These fits produce a density-weighted apparent ion temperature of $\langle T_{\mathrm{app}} \rangle = 1.55\pm0.24$\, keV}
    \label{fig:CycFits}
\end{figure}

Figure~\ref{fig:CycFits} shows the resulting fits for the cycloidal model. This model, which takes as inputs the known experimental parameters $V$, $B_z$, $a$, and $b$, is fit to the data by varying the ionization profile of the plasma.  Since the experimental voltage varied over the course of the OES exposure duration as $V=-33\pm5$kV, we show the resulting fits at three voltages.  The voltage decay during the measurement should narrow the distribution, which means these fits should be an underestimate of the instantaneous apparent ion temperature early in the plasma evolution.  

The fits do a reasonable job of capturing the overall width of the experimental data.  However, the fits have sharper features than are experimentally observed.  The fits are also unable to capture the fast ions opposing the plasma rotation (red shifted data for $y_{LOS}=15$mm and $30$mm and blue shifted data for $y_{LOS}=-22.5$mm and $-37.5$mm).  These differences lead to the high reduced $\chi^2$ and are most likely a result of the lack of velocity randomization (i.e. collisions) in this model.  From the fitted ionization profile, the density-weighted apparent ion temperature is calculated to be $\langle T_{\mathrm{app}} \rangle = 1.55\pm0.24$\,keV with the error bars set by fitting with different assumed voltages.

\subsection{The Application of the Rotating Gaussian Model to Data}
\begin{figure}
    \centering
    \includegraphics[width=0.95\linewidth]{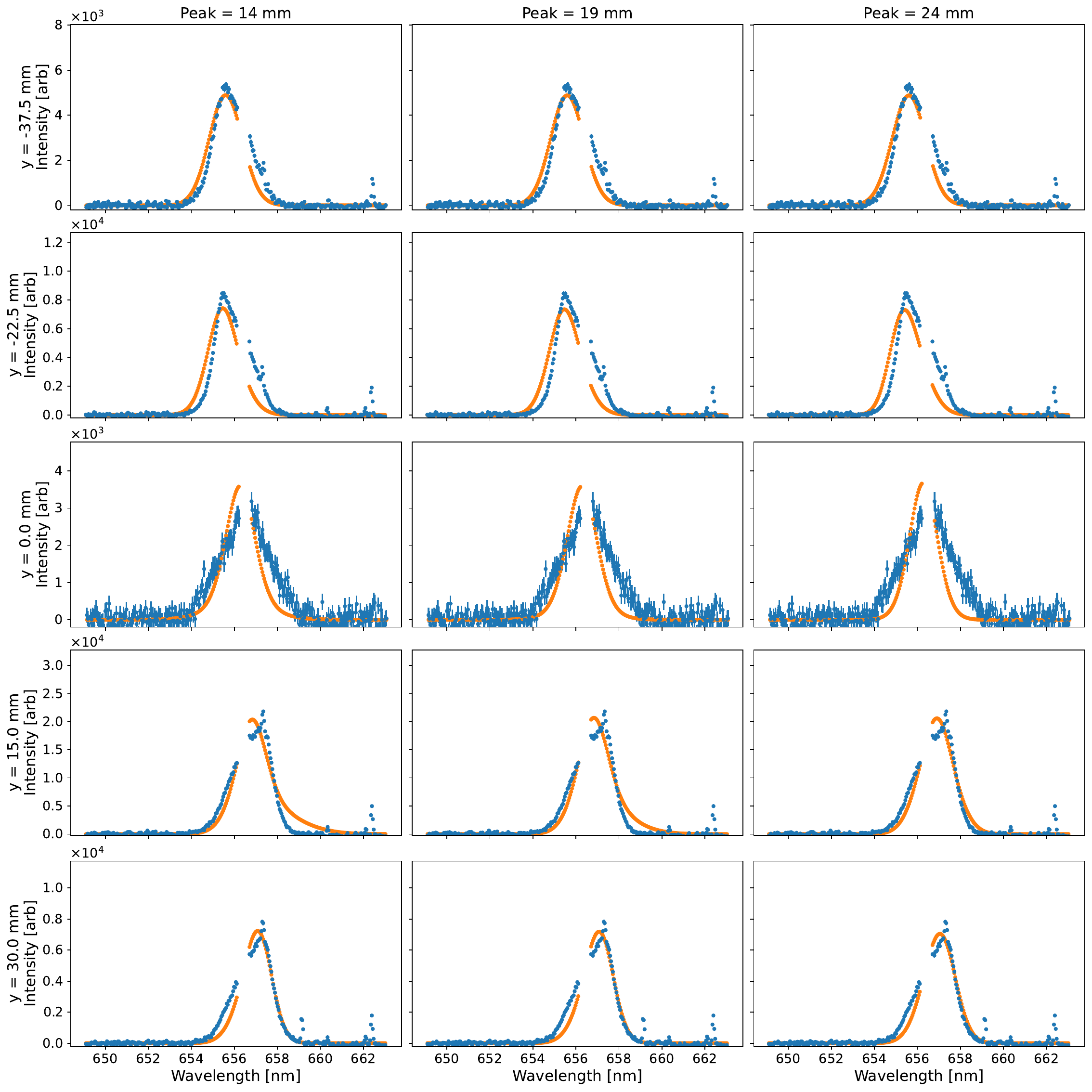}
    \caption{Measured $H_\alpha$ spectra (blue points) and best-fit model predictions (orange points) for the five OES viewing chords using the rotating Gaussian model, which assumes complete velocity-space relaxation to a drifting
    Maxwellian. The three different columns show separate fits using the same average voltage $V=-33$ kV, but with three different assumed linear peaked density profile with the peak density occurring at $r_{max}=(14,19,24)$ mm (Left, Middle, and Right Columns).  The measured spectra corresponds to a density-weighted apparent ion temperature of about $\langle T_{\mathrm{app}} \rangle = 1.40\pm0.43$\, keV representing the mean and standard deviation from nine fits with three different assumed profiles and three different voltages $V=-33\pm5$ kV.}
    \label{fig:RotFits}
\end{figure}

The application of the rotating Gaussian model to the data is shown in Fig.~\ref{fig:RotFits}. The apparent ion temperature profile is determined by fitting all LOS directly with $10$ temperature nodes used in the fitting and interpolating between these nodes.  A smoothness penalty is added to the cost function,
\begin{equation}
\eta \sum_k (T_{k+1} - 2T_k + T_{k-1})^2,\label{eqn:regularization}
\end{equation}
with $\eta=25$ to reduce unphysical oscillation in the apparent ion temperature.  This smoothing parameter has been chosen using the methodology previously discussed.  

The center of the Gaussian for each LOS is mainly determined by the density and rotation since it captures the LOS averaged rotational flow.  The linearly peaked density profile and $E\times B$ vacuum rotation capture this center fairly well. In Fig.~\ref{fig:RotFits}, we show the fits from the three different assumed linear density profiles shown in Fig.~\ref{fig:DenTempProf} with a peak density occurring at $r_{max}=(14,19,24)$mm.  The influence of the density profile is clearly seen in the $y_{LOS}=15$mm data.  As the peak density is moved closer to the cathode, where the $E\times B$ drift velocity is highest, the model has more rotational broadening, which is not captured in the experimental data.  

In general, the rotational broadening alone is unable to capture the observed width of the experimental spectra (blue). Significant additional thermal broadening of the spectrum is required to capture the data with the model (orange).  For this model, we are assuming a density profile that is qualitatively supported by radial dependence of the cold $H_\alpha$ intensity in Fig.~\ref{fig:RawCorr}.  We assume three different density profiles and three different voltages $V=-33\pm5$ kV to predict nine different density-weighted apparent ion temperatures.  These fits with the rotating Gaussian model give $\langle T_{\mathrm{app}} \rangle = 1.40\pm0.43$\, keV.

\subsection{Density and Temperature Profiles}
\begin{figure}
    \centering

    \begin{subfigure}{0.495\textwidth}
        \centering
        \includegraphics[width=\linewidth]{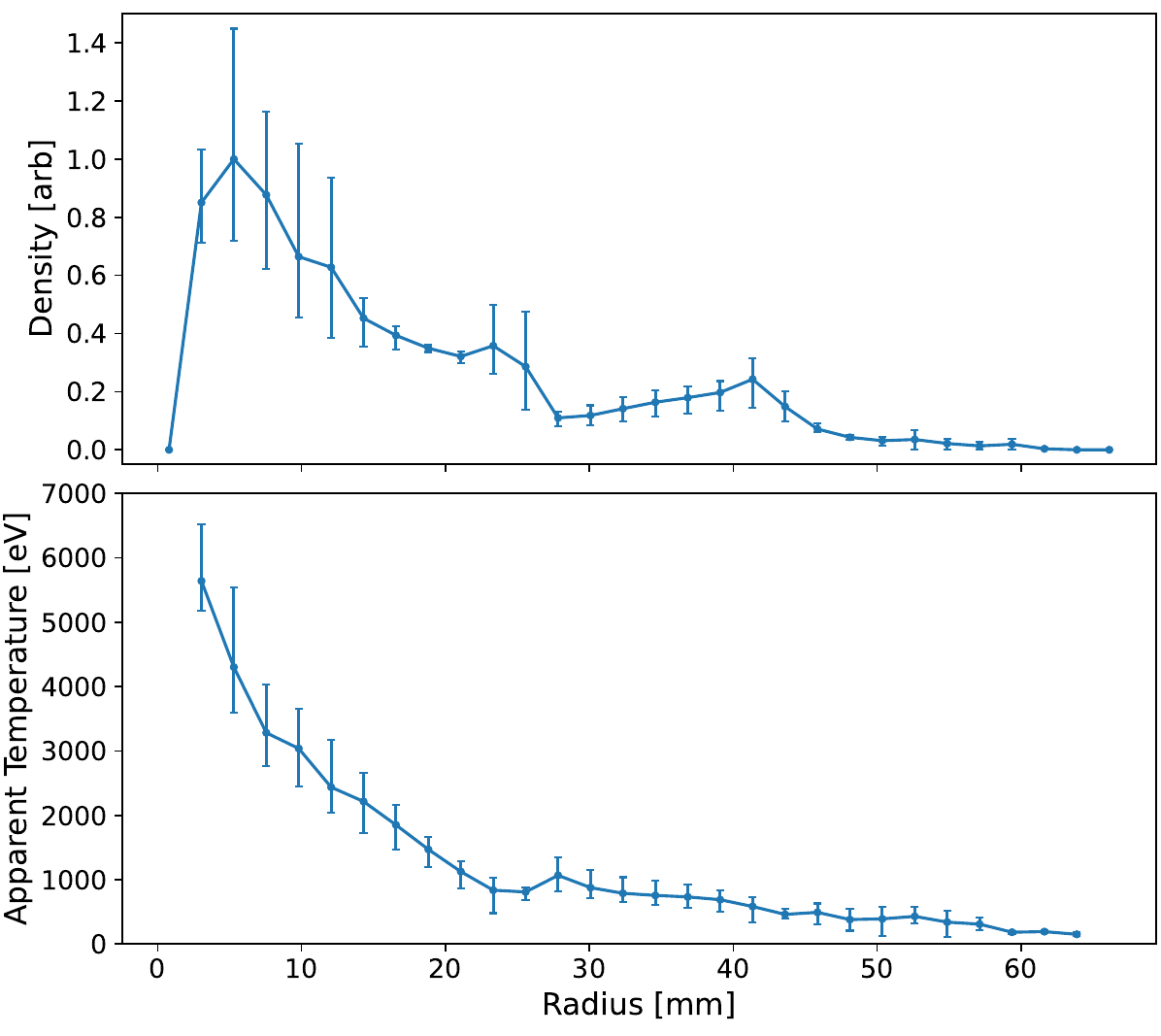}
    \end{subfigure}
    \hfill
    \begin{subfigure}{0.495\textwidth}
        \centering
        \includegraphics[width=\linewidth]{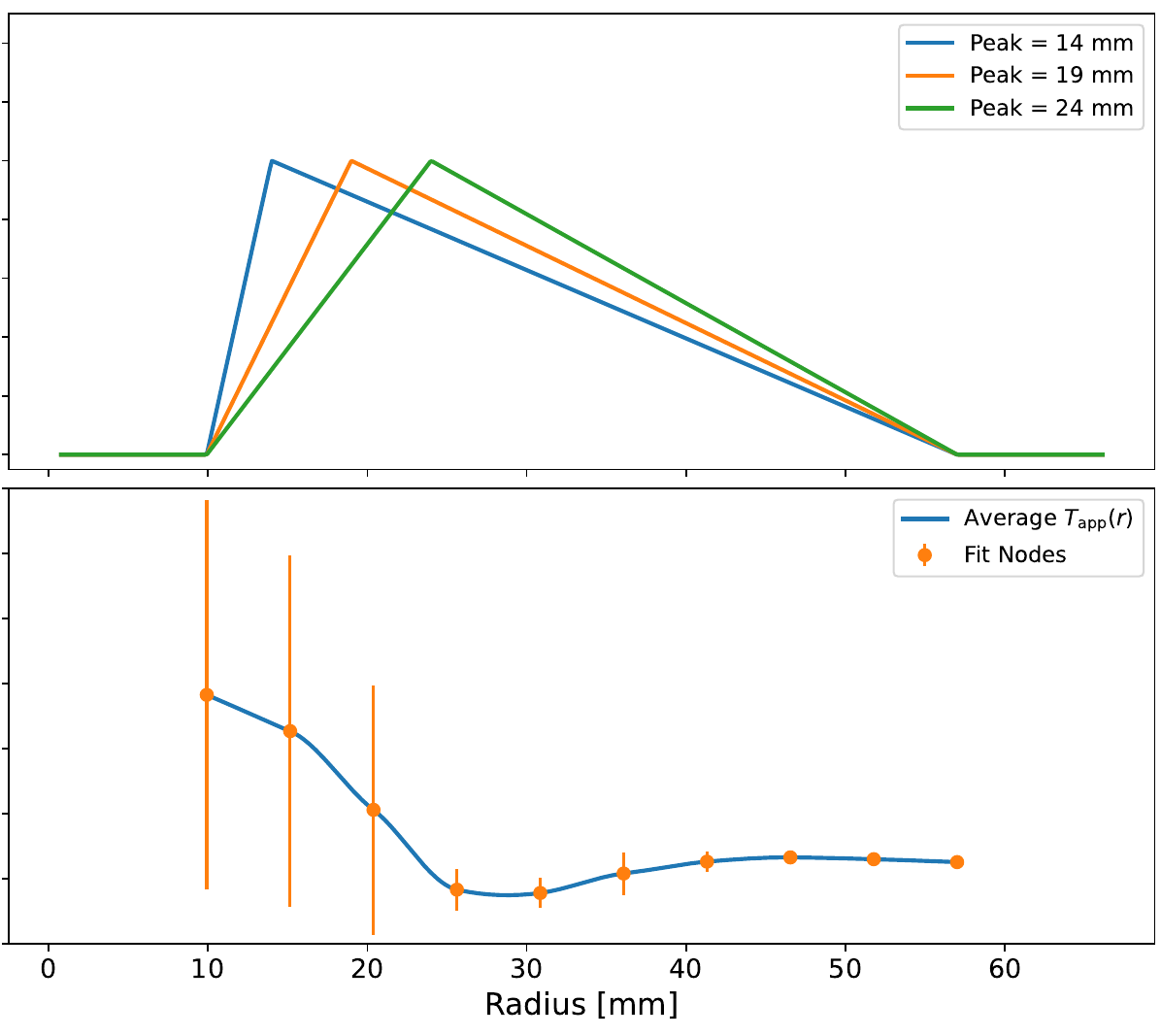}
    \end{subfigure}

    \caption{Inferred radial density and apparent ion temperature profiles. Left column: profiles obtained using the collisionless cycloidal model. Right column: profiles obtained using the rotating Gaussian model. In the rotating Gaussian analysis, the density is constrained to increase linearly from the first good flux surface at $r = 10$ mm to a peak at $r_{max}$, and then decrease linearly to zero at the last good flux surface $r = 57$ mm. The apparent ion temperature is treated as a free radial function and determined by minimizing the reduced $\chi^2$. Error bars represent the minimum and maximum from multiple simulations run with different applied cathode voltages $V=-33\pm5$ kV and the assumed density profiles shown for the rotating Gaussian model. Both models indicate density-weighted apparent ion temperatures above $1$ keV.}
    \label{fig:DenTempProf}
\end{figure}

Figure~\ref{fig:DenTempProf} shows the inferred density and temperature profiles when applying both models to the data.  For the cycloidal model (Left Column) the density and apparent ion temperature profiles are directly determined from the ionization profile that is the result of the $\chi^2$ fitting.  The peaked density near the cathode is consistent with the higher intensity cold $H_\alpha$ line experimentally observed in Fig.~\ref{fig:RawCorr}.  The error bars on these profiles are determined from fits using the average and extreme measured voltages during the exposure $V=-33\pm5$ kV.  The points are the average of these three fits and the error bars represent the minimum and maximum values.    

In contrast, for the rotating Gaussian model (Right Column of Fig.~\ref{fig:DenTempProf}), the density profiles have been assumed as the simplest profile that captures the previously discussed behavior of the cold $H_\alpha$ intensity.  The apparent ion temperature profile is a direct result of the fitting.  Here, we are showing the mean apparent ion temperature profile from the nine fits.  Error bars in the fitting nodes (orange) capture the standard deviation of these fits.  We find that the apparent ion temperature near the cathode varies substantially across these different fits.  In contrast, the uncertainty in the apparent ion temperature near the anode is small. 

The apparent ion temperature profiles from both fits have qualitatively similar behavior.  The cycloidal model is heavily constrained by the physics assumptions.  For this model, orbits near the cathode will have a larger variance and therefore a higher apparent ion temperature due to the higher electric field gradient.  This model provides a better fit to the spectra measured at $y_{LOS}=0$ mm.  In contrast, the rotating Gaussian model gives less weight to $y_{LOS}=0$ mm but provides better fits to the LOS closer to the anode.

\section{Summary and Path Forward}
\label{Sec:Sum}
OES measurements show evidence of ion rotation and Doppler broadening of fast charge exchange neutrals in Avalanche Energy's compact fusion device.  Two models of the LOS integrated distribution were presented that represent asymptotic limits in the ion collision rate.  The cycloidal model captured the width of the distribution, but contained sharper features than experimentally observed.  This model gives $\langle T_{\mathrm{app}} \rangle = 1.55\pm0.24$ keV.  A rotating Gaussian model assuming a linearly peaked density profile with vacuum $E\times B$ rotation captured the spectral variance of most of the LOS by fitting an apparent ion temperature profile and gives $\langle T_{\mathrm{app}} \rangle = 1.40\pm0.43$ keV.  These are consistent with each other and constrain the density-weighted apparent ion temperature to above $1$ keV.

Future work will focus on improving both the experimental measurements and the physical modeling used to infer the apparent ion temperature. In particular, future analysis will incorporate the energy dependence of the charge-exchange reaction rate coefficient directly into the model to better reconstruct the underlying ion velocity distribution and reduce systematic uncertainty in the inferred apparent ion temperatures. Shorter optical exposure durations and time-resolved measurements will also be implemented to reduce temporal averaging over the cathode voltage decay, which underestimates the instantaneous peak ion energies. Experimentally, future operation at higher applied voltages, stronger magnetic fields, and larger device radii will explore scaling toward higher apparent ion energies and improved confinement in next-generation centrifugal mirror devices.

\section*{Acknowledgments}
This research partially used resources of the National Energy Research Scientific Computing Center, a DOE Office of Science User Facility, supported by the Office of Science of the U.S. Department of Energy under Contract No. DE-AC02-05CH11231 using NERSC Award No. FES-ERCAP0036860. This research used the open-source particle-in-cell code WarpX. Primary WarpX contributors are with LBNL, LLNL, CEA-LIDYL, SLAC, DESY, CERN, Helion Energy, TAE Technologies, and Realta Fusion. We acknowledge all WarpX contributors.

\bibliographystyle{unsrt}
\bibliography{bib}

\end{document}